\documentclass{iaus}
\usepackage{graphicx}

\title[The GRB luminosity function] %% give here short title %%
{The Luminosity Function of Long Gamma-Ray Burst and their rate at $z\ge 6$}

\author[R. Salvaterra et al.]   %% give here short author list %%
{R.~Salvaterra$^1$,
S.~Campana$^1$, G.~Chincarini$^{1,2}$, T.R.~Choudhury$^3$, S.~Covino$^1$,
A.~Ferrara$^4$, S.~Gallerani$^5$, C.~Guidorzi$^1$,
\and G.~Tagliaferri$^1$}

\affiliation{$^1$ INAF, Osservatorio Astronomico di Brera, via E. Bianchi 46, I-23807 Merate (LC), Italy \\[\affilskip]
$^2$  Universit\`a degli Studi di Milano
Bicocca, Piazza della Scienza 3, I-20126 Milano, Italy\\[\affilskip]
$^3$ Institute of Astronomy, Madingley Road, Cambridge CB3 0HA, UK\\[\affilskip]
$^4$ SISSA/International School for Advanced Studies, Via Beirut 4,
I-34100 Trieste, Italy\\[\affilskip]
$^5$ Institute of Physics, E\"otv\"os University, P\'azm\'any  P. s. 1/A, 1117
Budapest, Hungary}

\pubyear{2008}
\volume{255}  %% insert here IAU Symposium No.
\pagerange{119--126}
\jname{Low-Metallicity Star Formation: From the First Stars to Dwarf Galaxies}
\editors{L.K. Hunt, S. Madden \& R. Schneider, eds.}

\def\lsim{\mathrel{\rlap{\lower 3pt\hbox{$\sim$}}\raise 2.0pt\hbox{$<$}}}
\def\gsim{\mathrel{\rlap{\lower 3pt\hbox{$\sim$}} \raise 2.0pt\hbox{$>$}}}

\def\Zsun{{\rm Z}_{\odot}}

\begin{document}

\maketitle

\begin{abstract}
We compute the luminosity function (LF) and the formation rate of long gamma 
ray bursts (GRBs) in three different
scenarios:  i) GRBs follow the cosmic star formation and their LF is constant 
in time; ii) GRBs follow the cosmic star formation but the LF varies with
redshift; iii) GRBs form preferentially in low--metallicity
environments. We then test model predictions against the {\it Swift} 3-year 
data, showing that scenario i) is robustly ruled out. Moreover, we show
that the number of bright GRBs detected by {\it Swift} suggests that GRBs 
should have experienced some sort of luminosity evolution with redshift, 
being more luminous in the past. Finally we propose to use the observations
of the afterglow spectrum of GRBs at $z\ge 5.5$  to constrain the
reionization history and we applied our method to the case of GRB~050904.
\end{abstract}

\firstsection % if your document starts with a section,
\section{Introduction}

Long Gamma Ray Bursts (GRBs) are powerful flashes of high--energy photons 
occurring at an average rate of a few per day throughout the Universe up to 
very high redshift (the current record is $z=6.29$). The energy source of a 
long GRB is believed to be associated to the collapse of the core of a 
massive star (see M\'esz\'aros 2006 for a review). One of the main goals of the
{\it Swift}  satellite (Gehrels et al. 2004) is to tackle the key issue
of the GRB luminosity function (LF). Unfortunately, although the number of
GRBs with good redshift determination has been largely increased by 
{\it Swift}, the sample is still too poor (and bias dominated) to allow a 
direct measurement of the LF. We use here the {\it Swift} 3-year data to 
constrain the GRB LF and its evolution (Salvaterra \& Chincarini 2007, 
Salvaterra et al. 2008b). Moreover, we show a possible use
of GRBs detected at $z\ge 5.5$ to study the history of reionization (Gallerani
et al. 2008).

\section{Model description}

The observed photon flux, $P$, in the energy band 
$E_{\rm min}<E<E_{\rm max}$, emitted by an isotropically radiating source 
at redshift $z$ is

\begin{equation}
P=\frac{(1+z)\int^{(1+z)E_{\rm max}}_{(1+z)E_{\rm min}} S(E) dE}{4\pi d_L^2(z)},
\end{equation}

\noindent
where $S(E)$ is the differential rest--frame photon luminosity of the source, 
and $d_L(z)$ is the luminosity distance. 
To describe the typical burst spectrum we adopt the
functional form proposed by Band et al. (1993), i.e. a broken power--law
with a low--energy spectral index $\alpha$, a high--energy spectral index
$\beta$, and a break energy $E_b=(\alpha-\beta)E_p/(2+\alpha)$, 
with $\alpha=-1$ and $\beta=-2.25$ (Preece et al. 2000).
%, Kaneko et al. 2006). 
In order to broadly estimate the peak energy of the spectrum, $E_p$, 
for a given isotropic--equivalent peak luminosity, 
$L=\int^{10000\,\rm{keV}}_{1\,\rm{keV}} E S(E)dE$, we assumed
the validity of the correlation between $E_p$ and $L$ (Yonetoku et al. 2004). 
%Ghirlanda et al. 2005).
%, which is basically a surrogate of the $E_p-E_{iso}$
%relation (Amati et al. 2002, Amati 2006).

%\begin{equation}
%E_p=337\mbox{ keV } \left(\frac{L}{2\times 10^{52}\mbox{ erg s}^{-1}}\right)^{0.49},
%\end{equation}

%\noindent
%where $L=\int^{10000\,\rm{keV}}_{1\,\rm{keV}} E S(E)dE$. Although the above
%correlation has an appreciable scatter, we will show that this does not 
%affect our results.

Given a normalized GRB LF, $\phi(L)$, the observed rate of 
bursts with $P_1<P<P_2$ is

\begin{equation}
\frac{dN}{dt}(P_1<P<P_2)=\int_0^{\infty} dz \frac{dV(z)}{dz}
\frac{\Delta \Omega_s}{4\pi} \frac{\Psi_{\rm GRB}(z)}{1+z} \int^{L(P_2,z)}_{L(P_1,z)} dL^\prime \phi(L^\prime),
\end{equation}

\noindent
where $dV(z)/dz$
%=4\pi c d_L^2(z)/[H(z)(1+z)^2]$ 
is the comoving volume 
element\footnote{We adopted the 'concordance' model values for the
cosmological parameters: $h=0.7$, $\Omega_m=0.3$, and $\Omega_\Lambda=0.7$.},
%and $H(z)=H_0 [\Omega_M (1+z)^3+\Omega_\Lambda+(1-\Omega_M-\Omega_\Lambda)(1+z)^2]^{1/2}$.
$\Delta \Omega_s$ is the solid angle covered on the sky by the survey,
and the factor $(1+z)^{-1}$ accounts for cosmological time dilation. 
Finally, $\Psi_{\rm GRB}(z)$ is the comoving burst formation rate. In this 
work, we assume that the GRB LF is described by a power law with an exponential
cut--off at low luminosities, i.e. $\phi(L) \propto 
(L/L_{\rm cut})^{-\xi} \exp (-L_{\rm cut}/L)$.

%\begin{equation}
%\phi(L) \propto \left(\frac{L}{L_{\rm cut}}\right)^{-\xi} \exp \left(-\frac{L_{\rm cut}}{L}\right).
%\end{equation}

We consider three different scenarios: {\bf i) no evolution model}, where
GRBs follow the cosmic star formation and their LF is constant 
in time; {\bf ii)  luminosity evolution model}, where GRBs follow the cosmic 
star formation but the LF varies with
redshift; {\bf iii) density evolution model}, where GRBs form preferentially in low--metallicity
environments.
% Therefore we consider here two families of models: (i) luminosity
%evolution models, where the cut--off luminosity in the GRB LF varies as 
%$L_{cut}=L_0 (1+z)^\delta$ and (ii) density evolution models, where GRBs
%form preferentially in galaxies with metallicity below a given threshold
%$Z_{th}$. 
In the first two cases, the GRB
formation rate is simply proportional to the global SFR, i.e. 
$\Psi_{\rm GRB}(z)=k_{\rm GRB} \Psi_\star(z)$. We use here the recent 
determination of the SFR
obtained by Hopkins \& Beacom (2006), slightly modified to match the
observed decline of the SFR with $(1+z)^{-3.3}$ at $z\gsim 5$ suggested
by recent deep--field data (Stark et al. 2006). For the luminosity evolution 
model, we also assume that the  cut--off luminosity in the GRB LF varies as 
$L_{cut}=L_0 (1+z)^\delta$. Finally, for density evolution case,  the GRB  
formation rate is obtained by convolving 
the observed SFR with the fraction of galaxies at redshift $z$ with
metallicity below $Z_{th}$ using the expression computed by Langer \& Norman 
(2006). In this scenario, $L_{cut}={\rm const}=L_0$.

\section{Results} 

\begin{figure}[b]
% \vspace*{-2.0 cm}
\begin{center}
 \includegraphics[width=2.5in]{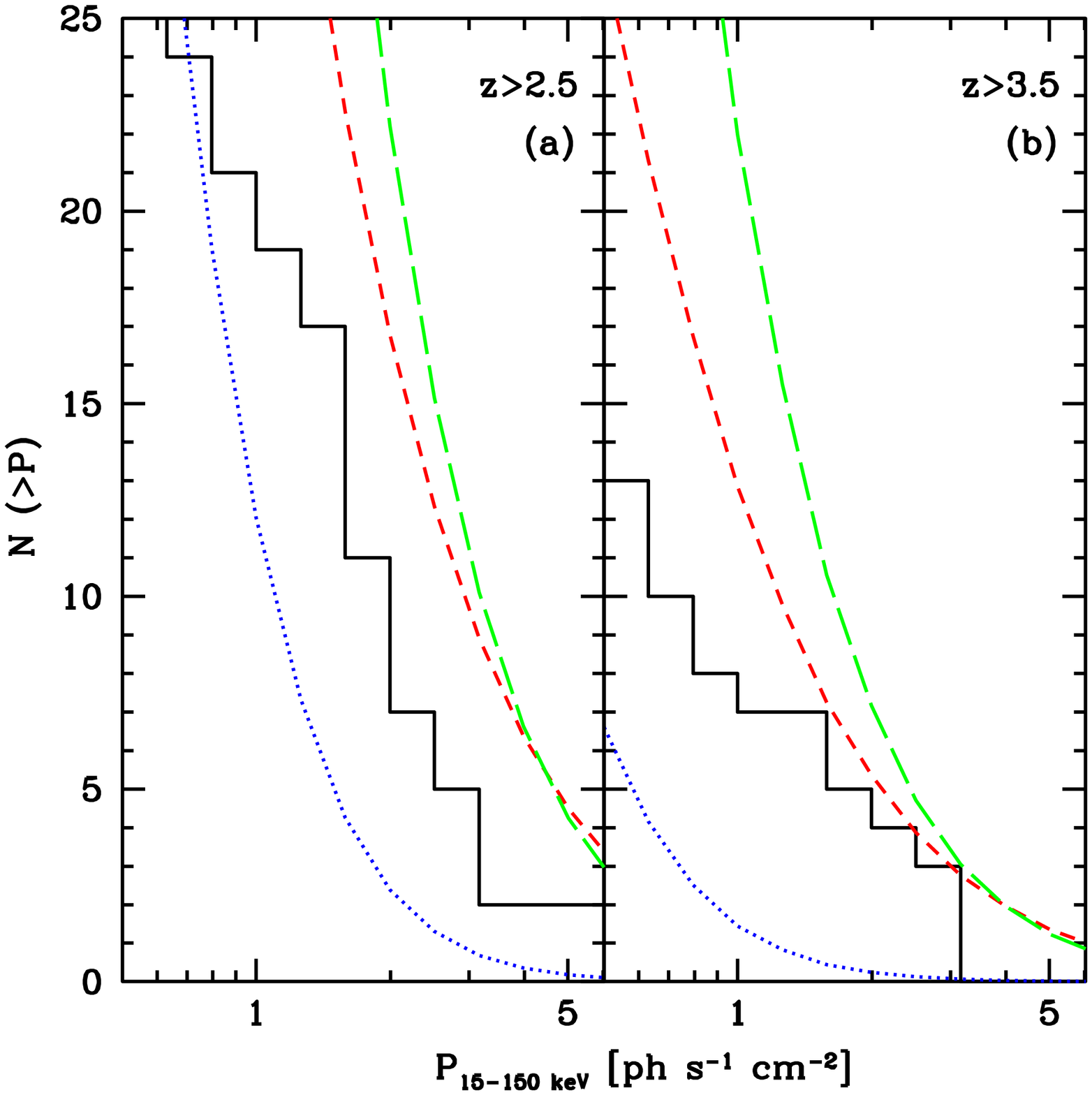} 
 \includegraphics[width=2.5in]{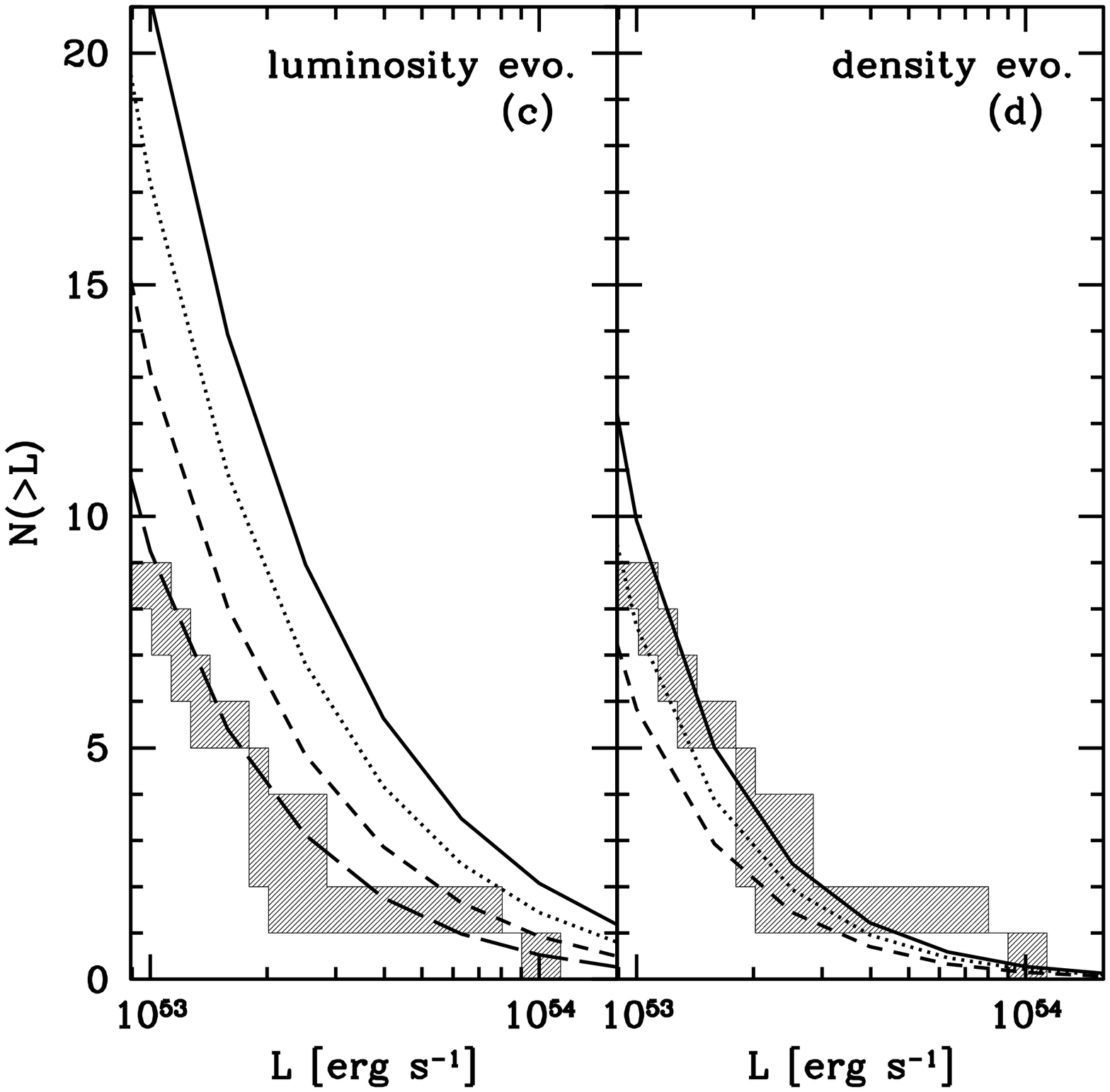} 
% \vspace*{-1.0 cm}
 \caption{{\bf Panels a \& b:} cumulative number of GRBs at $z>2.5$ (a) and
at $z>3.5$ (b) as a function of the photon flux $P$. Dotted line
refers to the no evolution model, short dashed to the luminosity evolution 
model ($\delta=1.5$) and long-dashed to the density evolution model 
($Z_{th}=0.1\;\Zsun$). 
The number of sources detected by {\it Swift} in three 
years is shown as solid histogram. Note that the observed detections 
are lower limits, since many high--$z$ GRBs can be missed by optical 
follow--up searches. A field of view of 1.4 sr for {\it Swift} is 
adopted.
{\bf Panels c \& d:} cumulative number of luminous GRBs detected by {\it Swift} in 
three years, shown with the histogram, as function of the 
isotropic equivalent peak luminosity, $L$. Shaded area takes into account  
the errors on the determination of $L$. Note that the data are to
be considered as lower limits of the real number of {\it Swift} 
detections.
For pure luminosity evolution models (panel c): solid 
line is for $\delta=3$, dotted line for $\delta=2.5$, short--dashed line for 
$\delta=2$, and long--dashed for $\delta=1.5$. 
For pure density evolution models (panel d): Solid line is for 
$Z_{th}=0.1\;\Zsun$, dotted line is for $Z_{th}=0.2\;\Zsun$, and short--dashed 
line is for $Z_{th}=0.3\;\Zsun$.}
   \label{fig1}
\end{center}
\end{figure}

The free parameters in our model are the GRB formation efficiency 
$k_{\rm GRB}$, the cut--off luminosity at $z=0$, $L_0$, and the LF power index
$\xi$. We optimized the value of these parameters by $\chi^2$
minimization over the observed differential number counts in the 50--300 keV
band of BATSE (Stern et al. 2000). We find that it is always possible to find
a good agreement between models and data. Moreover, we can reproduce also 
the 3--year differential peak flux count 
distribution in the 15-150 keV {\it Swift} band without changing the
best fit parameters
%using the same
%GRB LF and formation efficiency obtained by fitting the BATSE data
(Salvaterra \& Chincarini 2007).
We then check the resulting redshift distributions in the light of  the
{\it Swift} 3--year data, focusing on the large sample of GRBs detected 
at $z>2.5$ and $z>3.5$ (Fig.~1 panels a \& b). 
%\ref{fig1}).
The no evolution model is ruled out by the number of sure high-$z$ GRBs. 
This result is robust since does not depend on the
assumed SFR at high-$z$ nor on the faint--end of the GRB LF.
In conclusion, the existence of a large sample of bursts at
$z>2.5$ in the {\it Swift} 3-year data imply that 
GRBs have experienced some kind of evolution, being more luminous or
more common in the past (Salvaterra \& Chincarini 2007).

In order to discriminate between luminosity and density evolution models,
we compute the number of luminous GRBs, i.e. bursts with isotropic peak
luminosity $L\ge 10^{53}$ erg s$^{-1}$ in the 1-10000 keV band (Salvaterra et 
al. 2008b). We compare model predictions 
with the number of bright bursts detected by {\it Swift}. Conservatively, 
our data sample contains only bursts with a good redshift measurement 
and whose peak energy was well constrained by 
{\it Swift} itself or other satellites (such as HETE-2 or Konus-Wind).
We stress here that this number represents a lower limit on
the real number of bright GRBs detected, since some luminous bursts without
$z$ and/or $E_p$ can be present in the {\it Swift} catalog.
Results  for the pure luminosity (density) evolution models are plotted in the 
panel c (d) of Fig.~1. Data are shown with the histogram where
the shaded area takes into account errors on the determination of $L$.
We find that models involving pure luminosity evolution requires 
$\delta\gsim 1.5$ to reproduce the number of known bright GRBs. 
On the other hand, models in which GRB formation
is confined in low--metallicity environments fall short to account for the
observed bright GRBs for $Z_{th}>0.1$. 
Assuming $Z_{th}=0.1\;\Zsun$, the model reproduces the observed number of
bright GRBs, taking also into account the errors in the determination of $L$. 
This means that essentially all bright bursts present in the 3-year 
{\it Swift} catalog have a measured redshift and well constrained peak 
energy. So, although this model can not be discarded with high confidence, 
the available
data indicate the need of some evolution in the GRB LF even for such a low 
value of $Z_{th}$.
For $Z_{th}=0.3\;\Zsun$, as required by collapsar models (MacFadyen \& Woosley 
1999), only $\sim 6$ bursts with 
$L\ge 10^{53}$ erg s$^{-1}$ should have been detected in three year, largely 
underpredicting the number of {\it Swift} sure identifications.
Thus, pure density evolution models, where the GRB LF is constant with 
redshift, are ruled out by the number of bright GRBs. In conclusion, 
available data suggest that
GRBs have experienced some luminosity evolution with cosmic time.

\section{GRBs from the reionization epoch}

\begin{figure}[b]
% \vspace*{-2.0 cm}
\begin{center}
 \includegraphics[width=3.in]{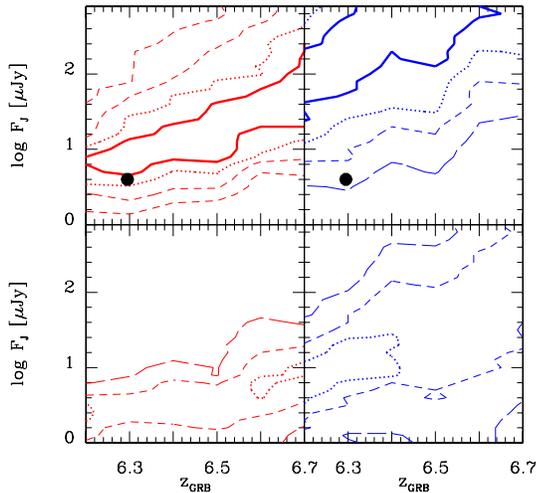} 
   \label{fig1}
\caption{Isocontours of the probability that the afterglow spectrum of J-band
flux $F_{\rm J}$ associated with a GRB at redshift $z_{\rm GRB}$, contains the
largest gap in the range 40--80~\AA~(top panels) and in the range 80--120 \AA~(bottom 
panel).  The left (right) panel shows the results for the ERM (LRM).
The isocontours correspond to probability of 15\% (long dashed line), 
30\% (short dashed line), 45\% (dotted line), and 60\% (solid line). The 
black point indicates the position in the $(z_{\rm GRB}, F_{\rm J})$ plane of GRB 050904.}
\end{center}
\end{figure}

We can now compute a robust lower limit on the  number of 
bursts detectable by {\it Swift} at very high-$z$. 
Assuming a trigger threshold $P\ge 0.4$ ph s$^{-1}$ cm$^{-2}$, at
least $\sim 5-10$\% of detected GRBs should lie at $z\ge 5$, with
$>1-3$ GRB yr$^{-1}$ at $z\ge 6$. 
These numbers double by lowering the {\it Swift} trigger threshold
by a factor of two (Salvaterra et al. 2008a).

High-$z$ GRBs are a useful and unique tool to study the Universe near and
beyond the reionization epoch. Gallerani et al. (2008) have studied the
possibility to constrain the reionization history using the statistics of
the dark portions (gaps) produced by intervening neutral hydrogen along the 
line of sight (LOS) in the afterglow spectra of GRB at $z\ge 5.5$. Two 
reionization models, both consistent with available observations of the
high-$z$ Universe, are considered: {\it (i) early reionization
model} (ERM) where $z_{reion}\sim 7$ and {\it (ii) late reionization
model} (LRM) where $z_{reion}\sim 6$.  
Suppose now that a GRB at redshift $z_{\rm GRB}$ is observed at a given flux 
level in the J band, $F_{\rm J}$. We can then ask what is the probability 
that the largest of the dark gaps in its afterglow spectrum is found within a 
given width range.  The results are shown in Fig.~2 for two different width
ranges; 
%40~\AA$\le W_{\rm max}\le 80$~\AA~(top panels) and 
%$80$~\AA$ \le W_{\rm max}\le 120$~\AA~(bottom panels);  
the left (right) panels refer to the ERM (LRM) case. The isocontours correspond to a probability of 15\%, 
30\%, 45\%, and 60\%. We find that the two models populate the $(z_{\rm GRB}, F_{\rm J})$ plane in a very
different way. In particular, for largest gaps in the 40--80 \AA~range, 
the highest probability is obtained for fainter afterglows in the ERM than  
for the LRM. 
%For example, at $z=6.5$ a probability $>60$\% is found for fluxes $6\lsim F_{\rm J}\lsim 40$ $\mu$Jy in the ERM, whereas similar 
%probabilities are found in the LRM only if $F_{\rm J}\gsim 160$
%$\mu$Jy. 
For largest gaps in the range 80--120 \AA,  the probability
is in general higher in the LRM with respect to the ERM. Note that, in
the ERM, only a few spectra should contain the largest gap in this 
range for $F_{\rm J} \gsim 10-40$ $\mu$Jy. 
Fig.~2 allows a straightforward comparison between data and model results. It 
is then natural to apply this procedure to GRB~050904 (black filled circle 
in Fig.~2). The probability to find the largest gap of 65~\AA~is $>45$\%
in the ERM, i.e. almost half of the LOS contains the largest 
gap in the range 40--80 \AA~for a burst with the redshift and flux of 
GRB~050904. Such probability drops for the LRM to $\sim 15$\%
clearly indicating that in this case the GRB~050904 observation represents 
a much rarer event. Although a large sample of high-$z$ GRBs is required 
before we conclude that a model in which reionization was complete at $z\sim 7$
is favored by the data, the discriminating power of the proposed method is 
already apparent. 

This kind of analysis requires high signal-to-noise, high resolution spectra 
of GRB afterglow spectra at $z\ge 5.5$ obtained with the largest ground
telescopes soon after the burst detection. To avoid 
wasting observing time, we developed a very effective strategy to spot reliable
$z\ge 5$ candidates on the basis of promptly available information provided
by {\it Swift} (Campana et al. 2007, Salvaterra et al. 2007). The selection
criteria adopted are: long burst observed durations ($T_{90}\gsim60$ s), faint $\gamma$-ray photon fluxes
($P\lsim 1$ ph s$^{-1}$ cm$^{-2}$), and no optical counterpart in the $V$ 
and bluer filters of UVOT ($V\gsim 20$).
We tested our selection procedure against the last $\sim 2$ years of 
{\it Swift} data showing that our method is very efficient and clean
(i.e. no low-$z$ interloper is present in the sample).

\section{Conclusions}

We have tested different formation and evolution scenarios for long GRB 
against the 3-year {\it Swift} dataset. We found that {\it Swift} data
strongly rule out models in which GRBs follow the cosmic star formation
and their LF is constant in time. In particular, the number of bright 
GRBs suggests that GRBs should have experienced some 
sort of luminosity evolution with cosmic time, being more luminous in the past.
Finally we have shown that GRBs at $z\ge 5.5$ can be use to constrain the
reionization history and we applied our method to the case of GRB~050904 at
$z=6.29$.

\end{document}